\documentclass[prl,twocolumn,showpacs,preprintnumbers,amsmath,amssymb]{revtex4}

\usepackage{amsmath,amssymb,amsfonts} 	
\usepackage[dvips]{graphicx}
\usepackage[latin1]{inputenc}
\usepackage{subfigure}
\begin{document}

\title{Extending the random-phase approximation for electronic correlation energies: The renormalized adiabatic local density approximation}

\author{Thomas Olsen}
\email{tolsen@fysik.dtu.dk}
\author{Kristian S. Thygesen}

\affiliation{Center for Atomic-Scale Materials Design,
	     Department of Physics, Technical University of Denmark,
	     DK--2800 Kongens Lyngby, Denmark}

\date{\today}

\begin{abstract}
The adiabatic connection fluctuation-dissipation theorem with the random phase approximation (RPA) has recently been applied with success to obtain correlation energies of a variety of chemical and solid state systems. The main merit of this approach is the improved description of dispersive forces while chemical bond strengths and absolute correlation energies are systematically underestimated. In this work we extend the RPA by including a parameter-free renormalized version of the adiabatic local density (ALDA) exchange-correlation kernel.
The renormalization consists of a (local) truncation of the ALDA kernel for wave vectors $q>2k_F$, which is found to yield excellent results for the homogeneous electron gas. In addition, the kernel significantly improves both the absolute correlation energies and atomization energies of small molecules over RPA and ALDA. The renormalization can be straightforwardly applied to other adiabatic local kernels.
\end{abstract}
\maketitle
Increasing computational resources has recently boosted a major interest in calculating electronic correlation energies from first principles using the adiabatic connection fluctuation-dissipation theorem (ACDF) \cite{lein, furche_voorhis, eshuis, dobson_wang00}. The computational cost for such methods is much higher than traditional correlation functionals in density functional theory, but has the great advantage that it includes non-local effects and does not rely on error cancellation between exchange and correlation. The RPA represents the simplest approach to ACDF calculations and has already been applied to broad range of electronic structure problems \cite{dobson_wang99, furche, harl08, harl09, harl10, lebegue, schimka}. While the non-locality of RPA makes it superior to semi-local functionals when dispersive interactions are important \cite{dobson_wang99, harl08, lebegue, olsen, mittendorfer, dobson_gould}, the accuracy for molecular atomization energies is comparable to that of the Perdew-Burke-Ernzerhof (PBE) functional \cite{furche, eshuis}. The relatively poor performance for atomization energies can be attributed to a deficient description of short-range correlation effects. Furthermore, total correlation energies are severely underestimated in RPA and an accurate description of energy differences is highly dependent on detailed error cancellation. A simple and intuitively appealing idea to remedy this problem was proposed by Yan et al. \cite{yan} (RPA+), however, the method does not seem to improve atomization energies although total correlation energies are much better described \cite{furche, harl08, eshuis}. From a perturbative point of view, RPA can be improved by including a screened second order exchange term (SOSEX) \cite{gruneis}, which exactly cancels the one-electron self-correlation energy of RPA, albeit, with a significant increase in computational cost. In addition, it has been shown that RPA results can be improved by explicitly including single excitation terms, which correct the use of non-selfconsistent input orbitals \cite{ren}.

From the point of view of time-dependent density functional theory (TDDFT), it is natural to try to improve the description of short-range correlation effects by extending RPA with an exchange-correlation kernel. For the homogeneous electron gas (HEG), this approach has been analyzed for a range of known exchange correlation kernels \cite{lein, fuchs} and led to the construction of new adiabatic kernels fitted to reproduce the HEG correlation energy \cite{dobson_wang00, jung}. So far, it seems that for accurate total energy calculations, the non-locality of exchange-correlation kernels is very important, whereas the frequency dependence is less critical. Moreover, the pair-distribution function derived from any local approximation for the exchange-correlation kernel exhibits an unphysical divergence at the origin \cite{furche_voorhis}. While correlation energies are still well defined, the divergence makes it very hard to converge correlation energies based on local kernels. Recently, a frequency dependent exact exchange kernel has been shown to produce accurate correlation energies for atoms and molecules \cite{hesselmann}. However, the computational cost of this approach is significantly larger than that of RPA and the method may not be directly applicable to periodic systems.

In this letter we derive a non-local exchange-correlation kernel, which does not contain any fitted parameters. The construction is based on a renormalization of the HEG correlation hole, which removes the divergence of the pair-distribution function and brings total correlation energies much closer to experimental values than both RPA and local approximations for the kernel.

From the adiabatic connection and fluctuation-dissipation theorem, it follows that the correlation energy of an electronic system can be written
\begin{align}
 E_c[n]=-\int_0^1d\lambda\int_0^\infty\frac{d\omega}{2\pi}\text{Tr}[v\chi^\lambda(i\omega)-v\chi^{KS}(i\omega)].
\end{align}
Here $\chi^{KS}$ is the exact Kohn-Sham response function and $\chi^\lambda$ is the interacting response function of a system where the electron-electron interaction $v$ has been replaced by $\lambda v$. Using TDDFT, one may express the interacting response function in terms of the Kohn-Sham response function as
\begin{align}\label{chi}
 \chi^\lambda = \chi^{KS} + \chi^{KS}f^\lambda_{Hxc}\chi^\lambda,
\end{align}
where $f^\lambda_{Hxc}=\lambda v+f^\lambda_{xc}$ is the Hartree-exchange-correlation kernel at coupling strength $\lambda$. The simplest approximation for $f^\lambda_{Hxc}$ is the random phase approximation where the exchange-correlation part is neglected. A natural next step is to include an adiabatic local approximation for the exchange correlation kernel. In particular, one could try the ALDA kernel 
\begin{align}\label{ALDA}
 f_{xc}^{ALDA}[n](\mathbf{r},\mathbf{r}') = \delta(\mathbf{r}-\mathbf{r}')f_{xc}^{ALDA}[n],
\end{align}
where $f_{xc}^{ALDA}[n]=\frac{d^2}{dn^2}(ne_{xc}^{HEG})\Big|_{n=n(\mathbf{r})}$.  In the following we will only consider the exchange part of the adiabatic kernel, since it has the simplifying property that $f^\lambda_x=\lambda f_x$. Additionally, we expect the effect of including a kernel in Eq. \eqref{chi} will be dominated by the exchange contributions. As it turns out, the kernel Eq. \eqref{ALDA} neither improves on total correlation energies \cite{lein} nor molecular atomization energies \cite{furche_voorhis} and is plagued by convergence problems related to the divergence of the pair-distribution function.

\begin{figure}[tb]
	\includegraphics[width=4.25 cm]{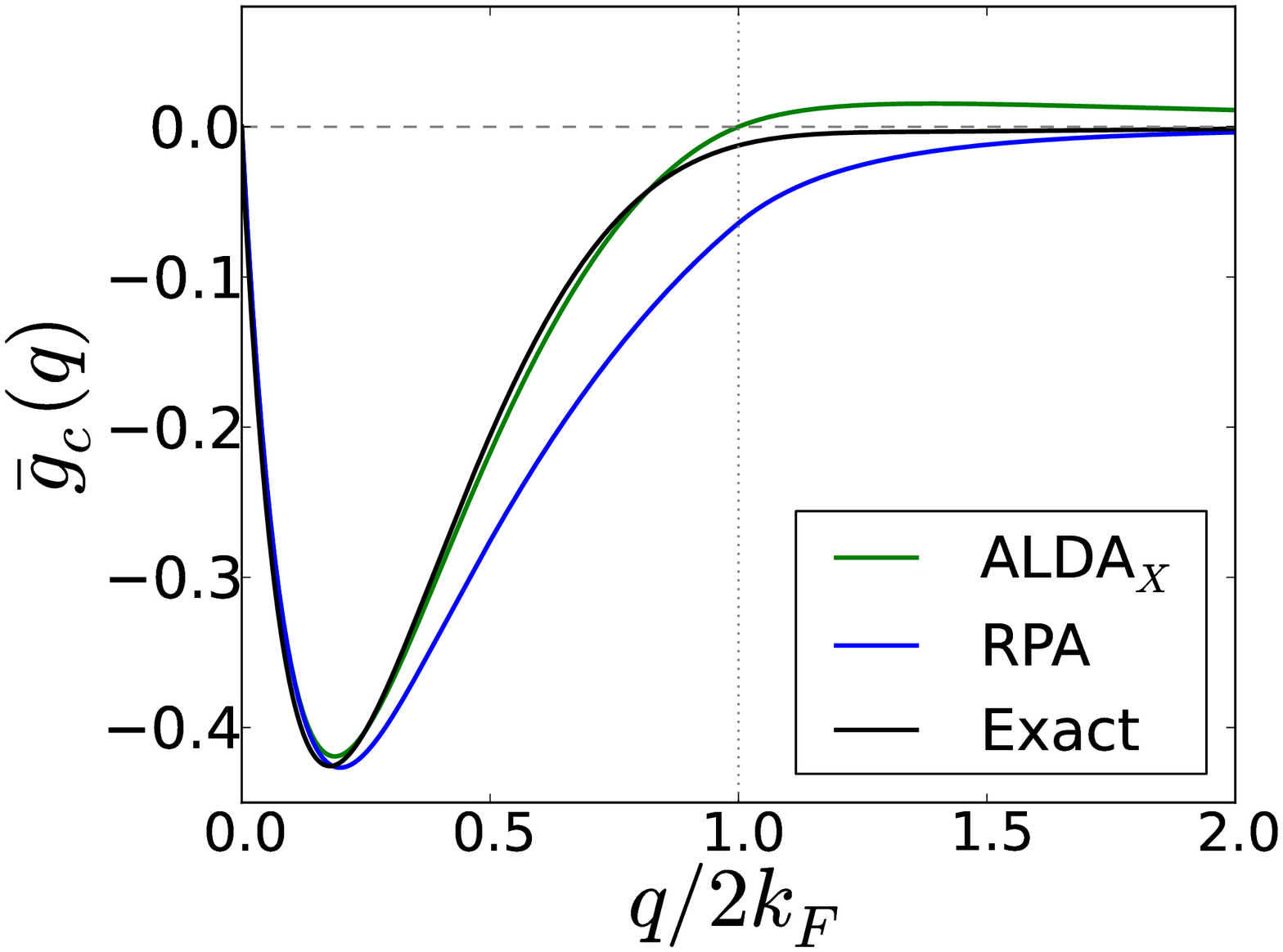} 
        \includegraphics[width=4.25 cm]{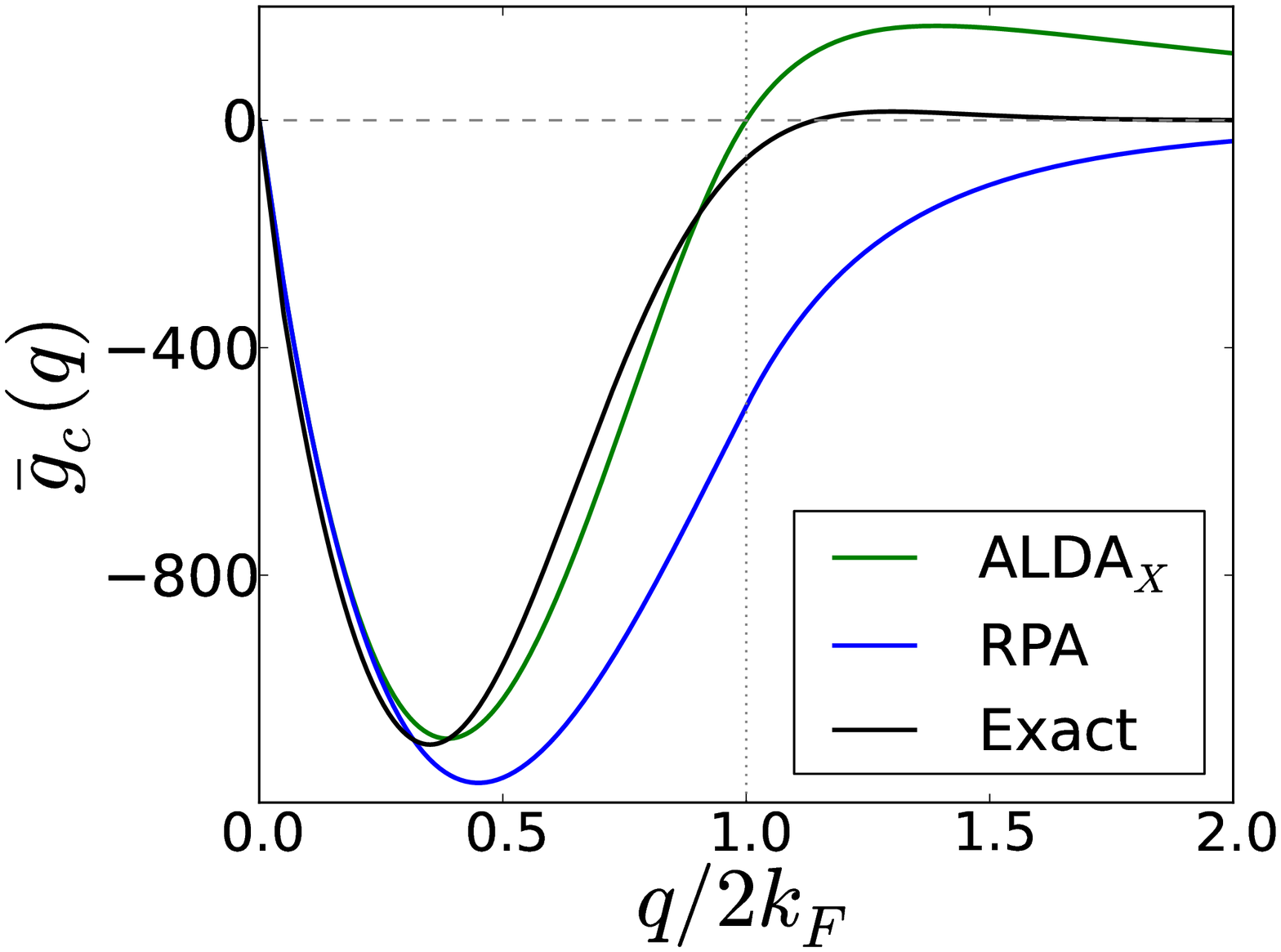}	
\caption{(color online). Fourier transform of the coupling constant averaged correlation hole for the homogeneous electrons gas. Left: $r_s=1$. Right: $r_s=10$.}
\label{fig:g_q}
\end{figure}

For the homogeneous electron gas the problem is naturally analyzed in reciprocal space where an accurate parametrization of the correlation hole is known \cite{perdew_wang, lein}. In Fig. \ref{fig:g_q} we show the exact coupling constant averaged correlation hole of the homogeneous electron gas and compare with RPA and ALDA$_X$ results. Whereas RPA underestimates the value at a large range of $q$-values, ALDA$_X$ gives a reasonable description at small $q$ but overestimates the value for $q>2k_F$. The ALDA$_X$ correlation hole becomes zero when $f_{Hx}^\lambda=0$, which happens exactly at $q=2k_F$. The divergence of the pair-distribution function originates from the slowly decaying tail at large $q$ where $f_{Hxc}$ is complete dominated by the $q$-independent $f_{x}^{ALDA}$ \cite{furche_voorhis}. The full ALDA correlation hole is very similar to the ALDA$_X$ correlation hole displayed here \cite{lein}.

\begin{figure}[tb]
	\includegraphics[width=6.0 cm]{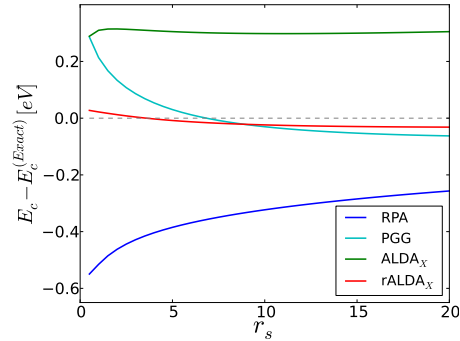} 
\caption{(color online). Correlation energy per electron of the homogeneous electron gas evaluated with different approximations for $f_{xc}$.}
\label{fig:heg_energy}
\end{figure}

The correlation energy is essentially given by the integral of the coupling contant averaged correlation hole. Despite the divergent pair-distribution function, the ALDA$_X$ correlation energy is well defined but converges very slowly due to the slow decay of $\bar g(q)$ at large $q$. From the shape of the correlation hole it is expected that RPA underestimates the correlation energy, while ALDA$_X$ overestimates it. Since the bad behavior of ALDA$_X$ primarily comes from large values of $q$ it is now tempting to introduce a renormalized ALDA$_X$ correlation energy obtained by cutting the $q$-integral at the zero point of $\bar g(q)$. The result is shown in Fig. \ref{fig:heg_energy} along with RPA, PGG \cite{pgg}, and ALDA$_X$ correlation energies. It is seen that the renormalized ALDA$_X$ gives a remarkable improvement compared to RPA, ALDA$_X$ and PGG. Except for the $r_s\rightarrow 0$ limit, it also performs better than the functionals proposed by Corradini et al \cite{corradini} and the static version of the Richardson-Ashcroft kernel \cite{richardson} (not shown), which were fitted to quantum Monte Carlo data and derived from many-body perturbation theory respectively \cite{lein}.

For the homogeneous electron gas, the cutoff is equivalent to using the Hartree-exchange-correlation kernel
\begin{align}
f_{Hxc}^{rALDA}[n](q)=\theta\Big(2k_F-q\Big)f_{Hx}^{ALDA}[n].
\end{align}
Fourier transforming this expression yields
\begin{align}\label{rALDA}
f_{Hxc}^{rALDA}[n](r)&=f_{x}^{rALDA}[n](r)+v^r[n](r),\\ 
f_{x}^{rALDA}[n](r)&=\frac{f^{ALDA}_{x}[n]}{2\pi^2r^3}\Big[\sin(2k_Fr)-2k_Fr\cos(2k_Fr)\Big],\notag\\
v^r[n](r)&=\frac{1}{r}\frac{2}{\pi}\int_0^{2k_Fr}\frac{\sin x}{x}dx.\notag
\end{align}
Since $k_F$ is related to the density, it is now straightforward to generalize this to inhomogeneous systems. We simply take $r\rightarrow|\mathbf{r}-\mathbf{r}'|$ and $k_F\rightarrow(3\pi^2\tilde n(\mathbf{r},\mathbf{r}'))^{1/3}$ with $\tilde n(\mathbf{r},\mathbf{r}')=(n(\mathbf{r})+n(\mathbf{r}'))/2$. Thus, we obtain a non-local functional with no free parameters by performing a simple local renormalization of the correlation hole. It can be regarded as an ALDA functional where the delta function in Eq. \eqref{ALDA} has acquired a density dependent broadening. At large separation it reduces to the pure Coulomb interaction and it is expected to retain the accurate description of van der Waals interactions characteristic of RPA. For example, in a jellium with $r_s=2.0$ two points separated by 5 {\AA} gives a renormalized interaction  $v^r[r_s=2](|\mathbf{r}-\mathbf{r}'|)=0.97v(|\mathbf{r}-\mathbf{r}'|)$ and the magnitude of the Coulomb part of the kernel is a factor of 30 larger than $f_x^{rALDA}$.

\begin{figure}[t]
	\includegraphics[width=7.0 cm]{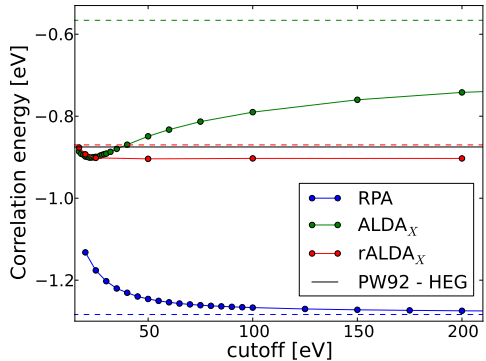} 
\caption{(color online). Correlation energy of the valence electron in Na evaluated with RPA, ALDA, and rALDA. The dashed lines show the values obtained with the functionals for the homogeneous electron gas using the average valence density of Na.}
\label{fig:Na}
\end{figure}
The renormalized ALDA functional has been implemented in the DFT code GPAW \cite{mortensen, gpaw-paper}, which uses the projector augmented wave (PAW) method \cite{blochl}. The response function is calculated in a plane wave basis set as described in Ref. \cite{jun}. The coupling constant integration is evaluated using 8 Gauss-Legendre points and the frequency integration is performed with 16 Gauss-Legendre points with the highest point situated at 800 $eV$. Since the kernel Eq. \eqref{rALDA} is only invariant under simultaneous lattice translation in $\mathbf{r}$ and $\mathbf{r}'$, its plane wave representation takes the form
\begin{align}
f^{rALDA}_{\mathbf{G}\mathbf{G}'}(\mathbf{q})=\frac{1}{V}\int_Vd\mathbf{r}\int_Vd\mathbf{r}'e^{-i\mathbf{G}\cdot\mathbf{r}}\tilde f(\mathbf{q};\mathbf{r},\mathbf{r}')e^{i\mathbf{G}'\cdot\mathbf{r}'},
\end{align}
where $\mathbf{G}$ and $\mathbf{G}'$ are reciprocal lattice vectors, $\mathbf{q}$ belongs to the first Brillouin zone, and
\begin{align}\label{f_tilde}
\tilde f(\mathbf{q};\mathbf{r},\mathbf{r}')=\frac{1}{N}\sum_{i,j}e^{i\mathbf{q}\cdot\mathbf{R}_{ij}}e^{-i\mathbf{q}\cdot(\mathbf{r}-\mathbf{r}')}f(\mathbf{r},\mathbf{r}'+\mathbf{R}_{ij}).
\end{align}
Here we have introduced the lattice point difference $\mathbf{R}_{ij}=\mathbf{R}_i-\mathbf{R}_j$ and the number of sampled unit cells $N$ (k-points). $\tilde f(\mathbf{q};\mathbf{r},\mathbf{r}')$ is thus periodic in both $\mathbf{r}$ and $\mathbf{r}'$ and $f^{rALDA}_{\mathbf{G}\mathbf{G}'}(\mathbf{q})$ should be converged by sampling a sufficient number of nearest neighbor unit cells. While the response function is calculated within the full PAW framework, it is not trivial to obtain the PAW corrections for a non-local functional and we use the bare ALDA$_X$ kernel to calculate contributions to the rALDA kernel from the augmentation spheres \cite{jun}.

As a first test of the functional for \textit{ab initio} applications, we have calculated the correlation energy of the valence electrons of bulk Na. We do not have a number for the exact value of the correlation energy but due to the delocalized nature of the valence electrons it is expected that the result should be close to the correlation energy of the homogeneous electron gas at the average valence density of Na. This is supported by the close agreement between the RPA correlation energy of Na and the homogeneous electron gas \cite{miyake}. We found the rALDA calculations to be converged when two nearest unit cells were included. The result is shown in Fig. \ref{fig:Na} as a function of plane wave cutoff energy along with the RPA and ALDA$_X$ results. As expected, RPA underestimates the correlation energy while ALDA$_X$ overestimates it. Again, one should note the slow convergence of the ALDA$_X$ calculation originating from the $q$-independent kernel. 
For plane wave implementations, an additional problem is posed by the divergens of $f_x^{ALDA}\sim n^{-2/3}$ at small densities. A particularly nice feature of the kernel \eqref{rALDA} is that the small density divergence of ALDA is regulated. For example, for small $r=|\mathbf{r}-\mathbf{r}'|$ one obtains
\begin{align}
f_{x}^{rALDA}[n](r)=4nf^{ALDA}_{x}[n],
\end{align}
whereas ALDA diverges. 

\begin{table}[b]
\begin{center}
\begin{tabular}{c|c|c|c|c|c|c}
      & LDA & PBE &  RPA & ALDA$_X$ & rALDA & Exact\\
	\hline
H     & -14 & -4  & -13 &   6  & -2   &   0 \\
H$_2$ & -59 & -27 & -51 & -16  & -28  & -26 \\
He    & -70 & -26 & -41 & -19  & -27  & -26
\end{tabular}
\end{center}
\caption{Correlation energies of H, H$_2$ and He evaluated with different functionals. Exact values are taken from Ref. \cite{lee}. All number are in kcal/mol.}
\label{tab:correlation}
\end{table}
The accuracy of molecular atomization energies by RPA is comparable to that of PBE, however, total correlation energies are typically severely underestimated. ALDA, on the other hand tend to overestimate total correlation energies by approximately the same amount. This is clearly seen for homogeneous systems displayed in Figs. \ref{fig:heg_energy} and \ref{fig:Na} and the trend is also observed for inhomogeneous systems. In Table \ref{tab:correlation} we show a few examples of atomic and molecular correlation energies calculated with the rALDA functional and compared with LDA, PBE, RPA, and ALDA results. The ACDF correlation energies were calculated in a 6x6x7 {\AA} unit cell. The RPA and rALDA results were calculated at increasing cutoffs up to $400\;eV$ and extrapolated to infinity. The ALDA$_X$ results were extrapolated from $1000\;eV$, but are still not well converged with respect to cutoff and represent a lower bound on the absolute ALDA$_X$ correlation energies. It is clear that the rALDA functional performs much better than both RPA and ALDA.

The significantly improved total correlation energies are a very nice feature of the rALDA kernel. However, most physical properties depend on energy differences and the kernel is not of much use if it does not perform at least as well as RPA for such quantities. In Tab. \ref{tab:atomization}, we display the atomization energies of a few simple molecules calculated with different methods. The RPA@LDA and RPA@PBE columns show the sum of Hartree-Fock and RPA energies evaluated at self-consistent LDA and PBE orbitals respectively. Whereas the Hartree-Fock term is nearly independent of input orbitals the RPA correlation energies show a significant dependence on the ground state functional. This dependence is unfortunate since there is no obvious choice for the set of input orbitals. In contrast, when an adiabatic approximation for the exchange-correlation kernel is used, a consistent choice is the ground state functional from which the kernel was derived \cite{furche_voorhis}. In the present case of ALDA and rALDA we thus only consider calculations on top of the LDA ground state. For these molecules the rALDA kernel is seen to underbind by a few kcal/mol (F$_2$ excepted) but is superior to the RPA and ALDA results.
\begin{table}[t]
\begin{center}
\begin{tabular}{c|c|c|c|c|c|c|c}
      & LDA  & PBE & RPA@LDA & RPA@PBE & ALDA & rALDA & Exp.\\
	\hline
H$_2$ & -113 & -105 & -109 & -109 (109) & -110 & -107 & -109 \\
N$_2$ & -268 & -244 & -225 & -224 (223) & -229 & -226 & -228 \\
O$_2$ & -174 & -144 & -103 & -112 (113) & -155 & -118 & -120 \\
CO    & -299 & -269 & -234 & -242 (244) & -287 & -253 & -259 \\
F$_2$ &  -78 &  -53 &  -13 &  -30 (30)  &  -74 & -39  &  -38 \\
HF    & -161 & -142 & -122 & -130 (133) & -157 & -136 & -141 \\
H$_2$O& -266 & -234 & -218 & -222 (223) & -249 & -225 & -233 \\
\hline
MAE   & 33 & 10.1 & 14.9 & 8.4 & 19 & 3.7 &  \\
\end{tabular}
\end{center}
\caption{Atomization energies of diatomic molecules. The ALDA values are taken from Ref. \cite{furche_voorhis} and experimental values (corrected for zero point vibrational energies) are taken from Ref. \cite{karton} Results in brackets are from Ref. \cite{furche}. All number are in kcal/mol. The bottom line shows the mean absolute error for this small test set.}
\label{tab:atomization}
\end{table}

The additional computational cost of calculating the kernel is insignificant compared to evaluating the non-interacting response function and inverting the Dyson equation. For a pure exchange kernel, it is possible to perform the coupling constant integration analytically, however, it involves an inversion of the non-interacting response function, which may become near singular at particular frequencies. The numerical coupling constant integration thus represents an additional computational cost compared to RPA calculations.

In summary, we have presented a new parameter free exchange kernel for total correlation energy calculations within the ACDF formalism. The kernel largely cancels the self-correlation energy of RPA and seems to perform better than both RPA and ALDA for molecular atomization energies as well as for simple metals. Allthough more benchmarking is needed, these preliminary results indicate that the rALDA functional is clearly superior to RPA. In contrast to RPA, the functional has the very nice feature that it provides a consistent choice of input orbitals beyond the Hartree approximation. Finally, it will be straightforward to extend the kernel to include ALDA correlation, which might be expected to improve results further, but we will leave this to future work. In fact, the renormalization method naturally generalizes to all semi-local adiabatic approximations, which all suffer from the same pathological behavior in their pair distribution functions and the present work just represents a single example of an entire class of renormalized adiabatic exchange-correlation kernels.

This work was supported by The Danish Council for Independent Research through the Sapere Aude program and the Danish Center for Scientific Computing.


\end{document}